\documentclass[twoside]{dis07}
\usepackage[latin1]{inputenc}
\usepackage[dvips]{graphicx,epsfig,color}
\usepackage{wrapfig,rotating}
\usepackage{amssymb,amsmath,array}

\begin{document}
\title{Summary of H1 Results on the Strong Coupling from Inclusive and Final States}

\author{Thomas Kluge
%
\vspace{.3cm}\\
%
DESY \\
Notkestr. 85, 22607 Hamburg - Germany
%
}

\maketitle

\begin{abstract}
Several determinations of the strong coupling from the H1 Collaboration are reviewed.
\end{abstract}

The following selection of results does not aim for completeness,
it emphasises the diversity of $\alpha_S$ determinations published by the H1 Collaboration
 and highlights those which are the most precise and have the largest range in scale. 

\begin{figure}
\begin{center}
\includegraphics[width=0.4\columnwidth]{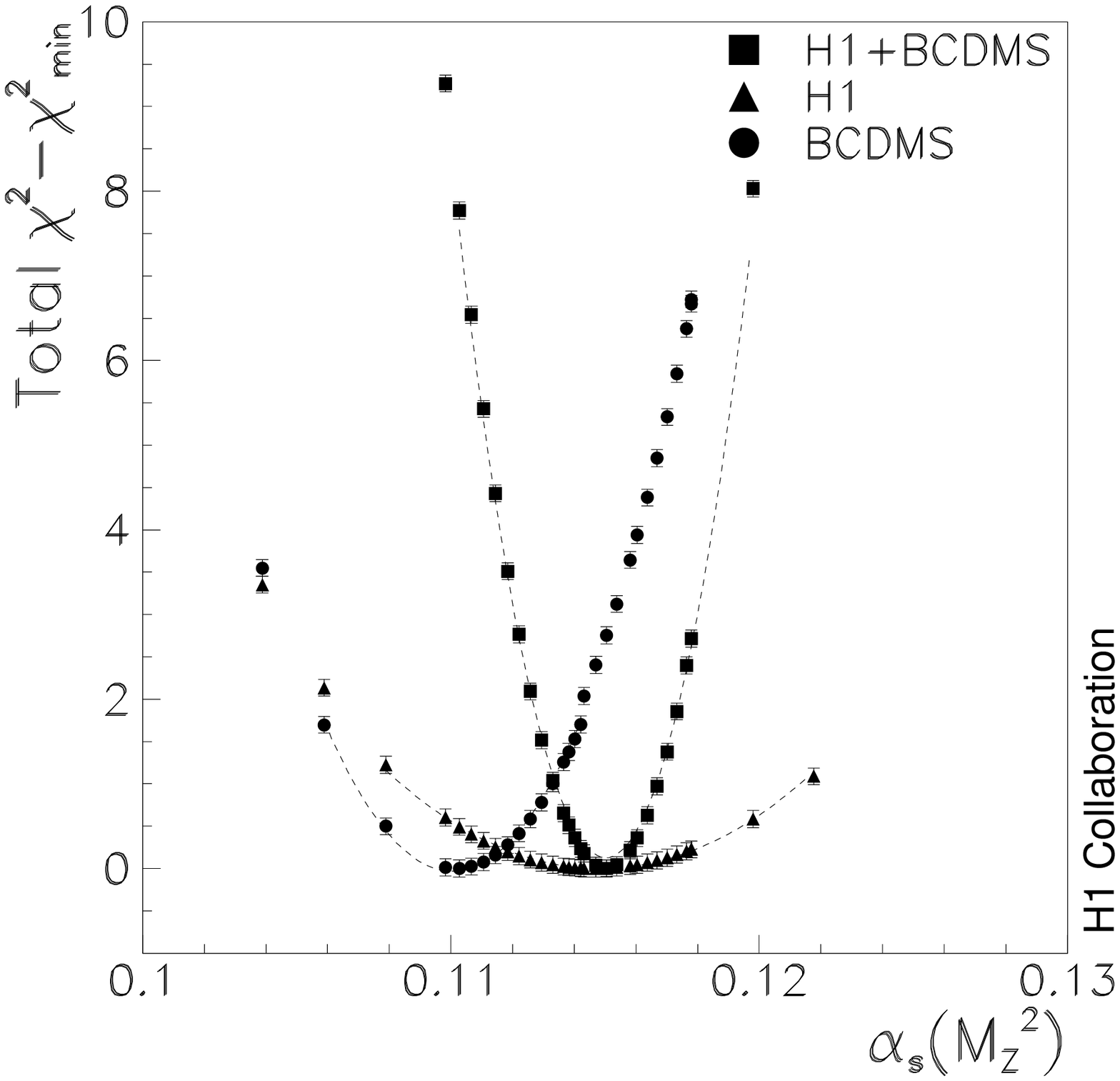}
\includegraphics[width=0.4\columnwidth]{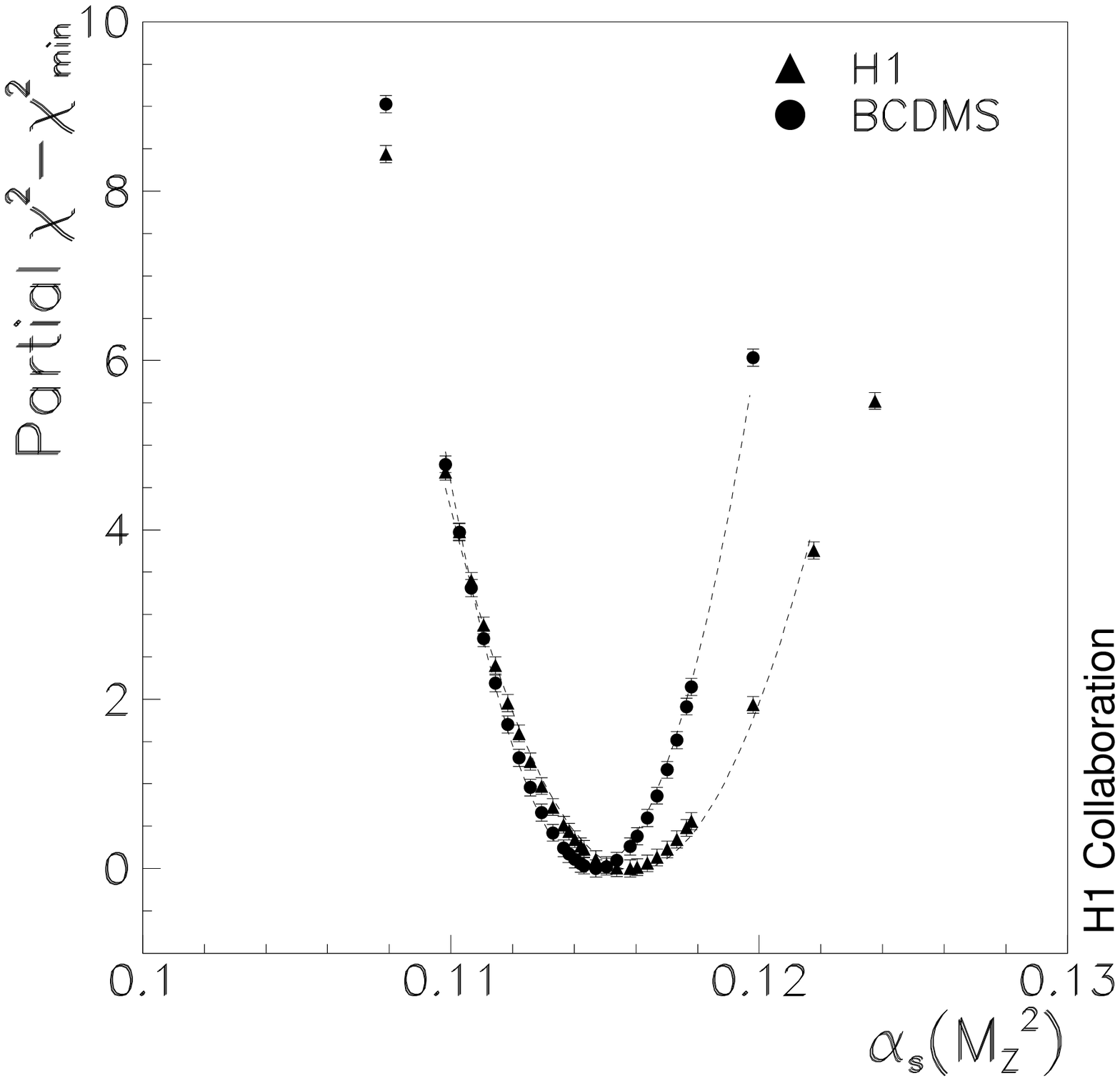}
\end{center}
  \caption{   Determination of the strong coupling constant $\alpha_S(M_Z)$ in NLO
   DGLAP QCD.
   a)~Total $\chi^2$ for fits to the H1 $ep$ and BCDMS $\mu p$ data 
   ($y_{\mu} > 0.3$) separately and for the  fit using
   data of the two experiments combined.
   b)~Partial $\chi^2$ contributions of the 
   H1 and BCDMS proton data in the fit to determine $\alpha_S$
   using both experiments.}
\label{fig:inclchi2}
\end{figure}

A simultaneous determination of the gluon distribution and of the strong coupling $\alpha_S(M_Z)$ is
obtained by combining low\,$x$ inclusive $e p$ scattering data of H1 with $\mu p$ scattering
data of the BCDMS collaboration at high\,$x$ \cite{Adloff:2000qk}.
A next-to-leading order (NLO) DGLAP QCD theory is fitted to the data.
The combined result is
$\alpha_S (M_Z) =0.1150 \pm 0.0017 (\rm exp.)
  ^{+0.0009}_{-0.0005}~(\rm model)\pm 0.005(\rm theo.)$.
The H1 data establish a strong rise of the gluon density towards low\,$x$, and consequently lead
to a depletion of the gluon at high\,$x$ in the fit.
Hence, the fit to the BCDMS data (at high\,$x$), when complemented with the H1 data, leads
 to a larger value of $\alpha_S(M_Z)$ than the fit to the BCDMS data alone (fig.~\ref{fig:inclchi2}a).
In the combined fit both data sets give a consistent and comparable contribution to the
error on $\alpha_S$ (fig.~\ref{fig:inclchi2}b).
The experimental error on the strong coupling is small compared to the theory error, which itself is 
 dominated by the uncertainty of the renormalisation scale.

\begin{figure}
\begin{center}
\includegraphics[width=0.4\columnwidth]{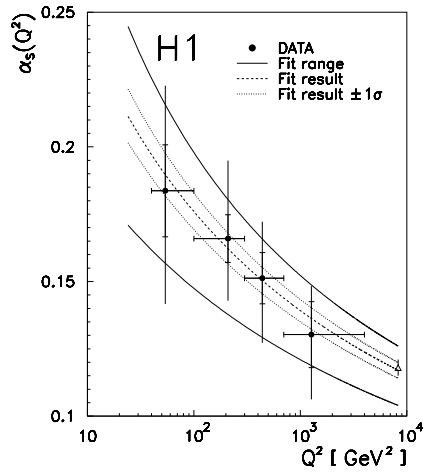}
\end{center}
  \caption{The $\alpha_S$ values derived from jet event rates as function of 
    $Q^2$ with statistical errors from the data and the correction 
    factors as inner error bars, and total errors as full error bars.
    The fit result (dashed line) and the 1 s.d.\ errors (dotted lines) 
    are based on the individual $\alpha_{S}(Q^2)$ values with 
    their statistical errors only. The solid lines represent the 
    obtained range for $\alpha_S(M_{Z})$ taking into 
    account the systematic uncertainties. 
    The open triangle at $Q^{2}= M_{Z}^2$  indicates a world average 
    $\alpha_S(M_{Z}) = 0.118 \pm 0.003$~\cite{PDG96}}.
\label{fig:jetrate}
\end{figure}

The probability of QCD gluon radiation in an $e p$ scattering event is governed by the strong coupling.
Hence, analyses of final states, e.g.\ of multi-jet events, yield constraints on $\alpha_S$.
In \cite{Adloff:1998kh} jets are defined using the modified JADE algorithm in DIS.
The observable is the (2+1) jet event rate, i.e.\ the relative fraction of 2-jet events in the sample as a function of $Q^2$.
This measurement is corrected for hadronisation effects to the parton level and compared to a perturbative
 calculation at NLO.
The result of a fit to the strong coupling is
$\alpha_S (M_Z) = 0.117 \pm 0.003 \,(\rm stat.) \, ^{+\,0.009}_{-\,0.013} \,(\rm sys.) \, + 0.006 \,(\rm jet\; algorithm)$.
Shown in fig.~\ref{fig:jetrate} is the result of the fit for four ranges in the scale $Q^2$,
 where the ``running'' of the strong coupling is demonstrated.
The systematic error is dominated by the uncertainty of the hadronic 
energy scale, the renormalisation scale dependence, and the dependence 
on the Monte Carlo model.

\begin{figure}
\begin{center}
\includegraphics[width=0.43\columnwidth]{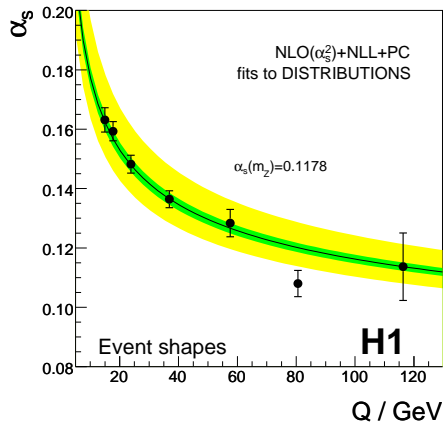}
\end{center}
  \caption{
    The strong coupling $\alpha_{S}$ as a function of the scale $Q$ from 
    an average of fits to differential event shape distributions.
    The individual results are compared with a common fit, the shaded areas 
    represent experimental and theoretical uncertainties.}
\label{fig:eventshape}
\end{figure}

Event shape variables are rather insensitive to hadronic energy scale uncertainties, 
since their definition uses only ratios of energies and momenta.
In \cite{Aktas:2005tz} results for five event shape variables in DIS are presented.
The observables are defined in the Breit frame of reference, where 
 the Born contribution generates no transverse momenta. 
Only particles in the ``current'' hemisphere of the Breit frame enter the definition of the event shape observables,
which eliminates sensitivity to the (non-perturbative) proton remnant.
Event shapes which are studied are two variants of thrust, the jet broadening, the jet mass and the $C$-parameter.
The theory prediction relies on NLO QCD and soft gluon resummation (NLL).
The rather large hadronisation effects are treated by Dokshitzer-Webber power corrections (PC).
Fits to the differential distributions of the event shape variables yield an average of
$  \alpha_S(M_Z) = 0.1198 \pm 0.0013\ ({\rm exp.})\ ^{+0.0056} _{-0.0043}\ ({\rm theo.})$.
The errors are dominated by the renormalisation scale uncertainty, which suggests
that missing higher order terms in the perturbative calculation are important.
Clear evidence of the asymptotic freedom of QCD as shown in fig.~\ref{fig:eventshape} is found when
fitting the results of individual ranges in the scale $Q$, and averaging over the event shapes. 
Note the large range in scale, from $15\,{\rm GeV}$ up to $116\,{\rm GeV}$, i.e.\ beyond the $Z$ mass, within one experiment and one analysis method.

\begin{figure}
\begin{center}
\includegraphics[width=0.7\columnwidth]{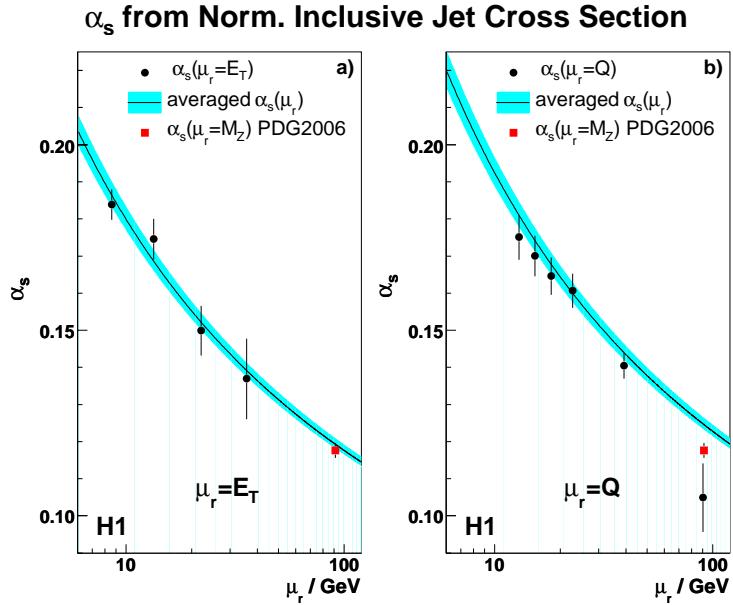}
\end{center}
  \caption{Determination of a) $\alpha_S(\mu_r=E_T)$ averaged over $Q^2$, and b) 
 $\alpha_S(\mu_r=Q)$ averaged over $E_T$.
The error bars denote the total experimental uncertainty for each data point.
The solid curve shows the result of evolving $\alpha_S(M_Z)$ averaged from all $Q^2$ and $E_T$ regions, 
 with the band denoting the total experimental uncertainty.
The world average from PDG is also shown.}
\label{fig:normjets}
\end{figure}

Analyses of inclusive jet cross sections in DIS and at not too small jet transverse energies
provide $\alpha_S$ determinations of high precision, because of only moderate higher order perturbative 
 contributions and small hadronisation corrections.
In a recent publication~\cite{Aktas:2007pb} double differential inclusive jet cross sections are presented
 as a function of $Q^2$ and of $E_T$ of the jets in the Breit frame.
The precision of QCD fits can be improved by using the {\it normalised} jet cross section, 
i.e.\ the ratio of the jet and the NC DIS cross section.
This observable benefits from a partial cancellation of experimental and theoretical uncertainties.
The measurements are found to be well described by
calculations at next-to-leading order in perturbative QCD, corrected for hadronisation effects.
A fit of the strong coupling yields
$\alpha_S(M_Z) = 0.1193 ~\pm 0.0014\,\mathrm{(exp.)}~ ^{+0.0047}_{-0.0030}\,\mathrm{(theo.)}~ \pm 0.0016\,\mathrm{(pdf)}$.
Fig.~\ref{fig:normjets} shows the results as function of $E_T$ and  $Q$
(and using $E_T$ and  $Q$ as renormalisation scale, respectively).

\begin{figure}
\begin{center}
\includegraphics[width=0.5\columnwidth]{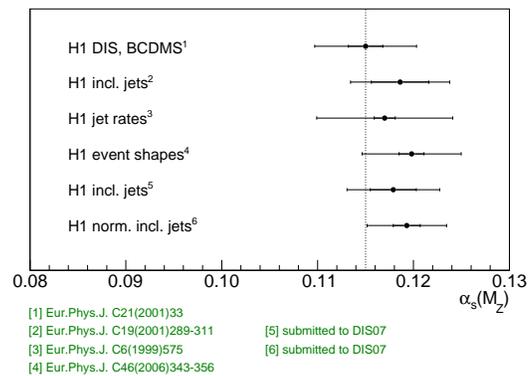}
\end{center}
\caption{Selection of $\alpha_S$ determinations at H1, presented with experimental errors (inner error bars) and  total errors (outer error bars).
}\label{fig:alphas_Summary}
\end{figure}

A comparison of the $\alpha_S(M_Z)$ determinations is shown on fig.~\ref{fig:alphas_Summary}.
The results are comparable with each other and with the world averages.
In the future, improvements in the precision are expected on the experimental side
 due to the large HERA~II dataset and on the theoretical side from NNLO calculations.


\begin{footnotesize}



\end{footnotesize}


\end{document}